\documentclass[%
reprint,
superscriptaddress,
 amsmath,amssymb,
]{revtex4-2}

\usepackage{graphicx}% Include figure files
\usepackage{dcolumn}% Align table columns on decimal point
\usepackage{bm}% bold math
\usepackage[table,x11names]{xcolor}
\usepackage[draft,inline,nomargin]{fixme}
\usepackage{comment}
\usepackage[normalem]{ulem}

\usepackage{booktabs}
\usepackage[margin=1in]{geometry}
\usepackage{makecell}
\usepackage{float}
\fxsetup{theme=color,mode=multiuser}

\FXRegisterAuthor{mh}{amh}{\color{blue}MH}
\FXRegisterAuthor{bh}{abh}{\color{red}BH}

\makeatletter
\renewcommand{\fnum@figure}{Fig.~\thefigure}
\makeatother

\begin{document}

\preprint{APS/123-QED}

\title{Unveiling Three-Dimensional Skyrmion Transitions Through Vortices and Monopoles}
%Tiwsts turns and tiny tornadoes: Exploring 3D skyrmions, vortices, and monopoles, in spintronics
%skyrmioniums and merons through bound vortices and monopoles
%through teh coupling of vortices and monopoles

\author{M. E. Henderson}
\email{hendersonme@ornl.gov}
\affiliation{Oak Ridge National Laboratory, Oak Ridge, TN 37831, USA}
\affiliation{Institute for Quantum Computing, University of Waterloo, Waterloo, ON, Canada, N2L3G1}
\affiliation{Department of Physics \& Astronomy, University of Waterloo,
  Waterloo, ON, Canada, N2L3G1}
%\noaffiliation

\author{D. Kurebayashi}
\affiliation{School of Physics, The University of New South Wales, Sydney 2052, Australia}

\author{B. Heacock}

\affiliation{National Institute of Standards and Technology, Gaithersburg, Maryland 20899, USA}

\author{W. Chen}
\affiliation{National Institute of Standards and Technology, Gaithersburg, Maryland 20899, USA}

\author{C. W. Clark}
\affiliation{National Institute of Standards and Technology, Gaithersburg, Maryland 20899, USA}

\author{D. G. Cory}
\affiliation{Institute for Quantum Computing, University of Waterloo, Waterloo, ON, Canada, N2L3G1}
\affiliation{Department of Chemistry, University of Waterloo, Waterloo, ON, Canada, N2L3G1}

\author{D. Sarenac}
\affiliation{Department of Physics, University at Buffalo, State University of New York, Buffalo, New York 14260, USA}

%\author{D. Sarenac}
%\affiliation{Institute for Quantum Computing, University of Waterloo, Waterloo, ON, Canada, N2L3G1}

\author{S. Watson}
\affiliation{National Institute of Standards and Technology, Gaithersburg, Maryland 20899, USA}

\author{J. S. White}
\affiliation{Laboratory for Neutron Scattering and Imaging, PSI Center for Neutron and Muon Sciences, Villigen, Switzerland}

\author{L. DeBeer-Schmitt}
\email{debeerschmlm@ornl.gov}
\affiliation{Oak Ridge National Laboratory, Oak Ridge, TN 37831, USA}

\author{O. A. Tretiakov}
\email{o.tretiakov@ unsw.edu.au}
\affiliation{School of Physics, The University of New South Wales, Sydney 2052, Australia}

\author{D. A. Pushin}
\email{dmitry.pushin@uwaterloo.ca}
\affiliation{Institute for Quantum Computing, University of Waterloo, Waterloo, ON, Canada, N2L3G1}
\affiliation{Department of Physics \& Astronomy, University of Waterloo,
 Waterloo, ON, Canada, N2L3G1}

\date{\today}
%TC:ignore
\begin{abstract}
Magnetic skyrmions represent vortex-like spin configurations that provide a robust platform for next-generation spintronic technologies. Although they are often treated as two-dimensional objects with integer topological charge, their extension into three-dimensional strings realizes other composite structures with unique device functionalities that transcend planar frameworks. Unfortunately, a lack of bulk probes has failed to realize such higher-dimensional topological structures and their implementations. Here, we report the first experimental visualization of three-dimensional topological $Q = 0$ skyrmion structures using neutron scattering tomography techniques across the equilibrium phase of a Co$_8$Zn$_8$Mn$_4$ sample. Disordered skyrmion states reveal metastable skyrmioniums and composite topological objects novel to bulk systems. Vortex-antivortex lattices mediate changes in topology, with unprecedented transition pathways via a coupling of merons and monopoles. The present realization of bulk Q = 0 quasiparticles and meron-mediated dynamics paves the way for higher-dimensional spintronic frameworks through multi-bit encoding architectures, unidirectional transport schemes, and monopole-mediated controls.

\end{abstract}

\maketitle

The evolution of topology with increasing dimensionality mediates a wide variety of exotic and hierarchical states, from singular defects \cite{kleman2006topological,chuang1991cosmology,kleman2008disclinations} to localized textures and superstructures \cite{al2001skyrmions,zheng2023hopfion,gobel2021beyond}. Fundamental to these objects is the action of vorticity, which characterizes the in-plane curling of a vector field around a core \cite{wachowiak2002direct}. In quantum magnets, this curling manifests across the spin degree of freedom, generating a diverse collection of non-collinear nanometric magnetic states with some characteristic topology. Integration of the magnetic vorticity over a closed 2D surface \cite{donnelly2021experimental} quantifies a topological charge $Q$, which counts the number of times the normalized spin field $\bm{n(r)}$ wraps the unit sphere: 
\begin{equation}
   Q = \frac{1}{4\pi} \int 
   \bm{n} \cdot \left( \frac{\partial \bm{n}}{\partial x}
   \times \frac{\partial \bm{n}}{\partial y} \right)
   \, dx\, dy .
   \label{eqn_top_charge}
\end{equation}

Values of $Q$ can vary arbitrarily and continuously above and below zero, spanning smoothly varying vortex-like spin configurations, such as integer $Q= \pm 1$ skyrmions \cite{Yu2010real,milde2013unwinding} and their half-integer $Q= \pm \frac{1}{2}$ meron composites \cite{yu2018transformation,wang2021meron,gao2019creation}, to abruptly vanishing singular magnetization distributions, such as Bloch points \cite{hermosa2023bloch,yasin2024bloch}. For clarity, we will refer to states with fractional topological charges, distinct from 1 and 1/2, as fractional vortices, while the term meron is reserved exclusively for half-skyrmion constituents possessing $|Q| = \frac{1}{2}$ \cite{tretiakov2007vortices}. Magnetic skyrmions constitute a unique combination of these topological states, serving as real-space sources of Berry curvature with defect mediated topologies \cite{schutte2014dynamics,Schulz2012emergent}. This collection of properties enables novel transport phenomena and emergent electrodynamics, inspiring a new generation of ultra-low power, high-density skyrmion-based spintronic devices \cite{Zhang2015magnetic,Fert2017magnetic}.

  \begin{figure*}[t]
\includegraphics[width =\textwidth]{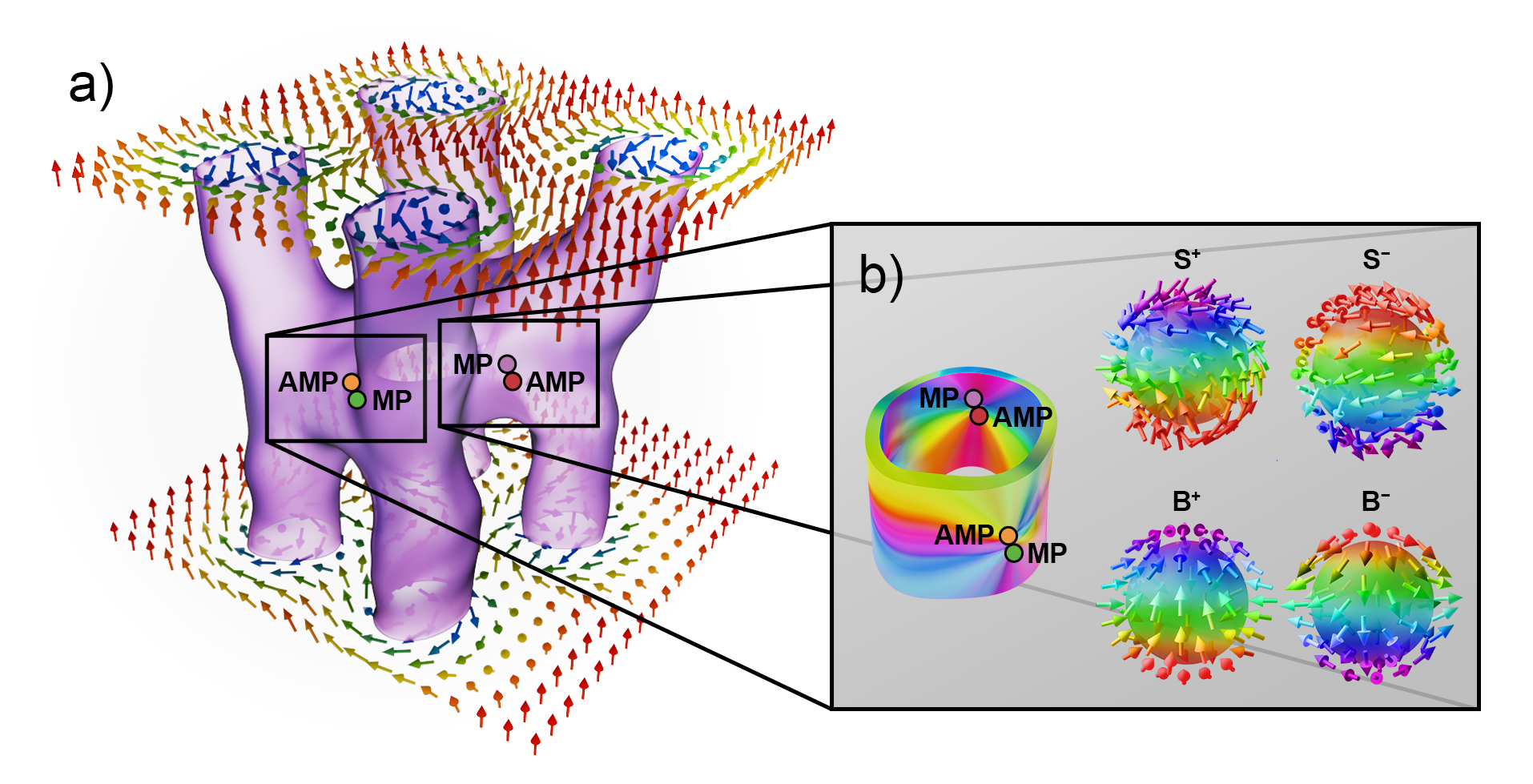}
\caption{Illustration of possible exotic skyrmionium stabilization pathway through the merging and separation of 4 individual skyrmion tubes. Coupled monopoles (MP) and antimonopoles (AMP) are shown at topological transition points, highlighting their corresponding Branching (B) and Segmenting (S) spin defects in the inset.         
}
\label{fig:overview}
\end{figure*}

Commonly visualized as a planar sheet in two dimensions \cite{Yu2010real}, in three dimensions the skyrmion cross-section extends into a string which can be interrupted by a singular Bloch point defect along its depth \cite{milde2013unwinding,henderson2023three,wolf2022unveiling,gobel2019magnetic}. In real space these defects resemble spin hedgehogs \cite{yasin2024bloch,Yu2020real}, while in momentum space these defects correspond to monopoles (MPs) and antimonopoles (AMPs) of the emergent magnetic field \cite{Yu2020real,Birch2021Topological,henderson2023three,jin2023evolution,yasin2024bloch}. Here, the gauge invariance of the skyrmion Berry curvature associates a quantized emergent flux with each skyrmion tube through a scalar spin chirality, coupling real-space magnetic defects with emergent MPs and AMPs \cite{nagaosa2012gauge,everschor2014real}. Skyrmion topological transitions are therefore thought to be dictated via the three-dimensional proliferation of emergent MPs/AMPs which nucleate/annihilate skyrmions uniquely at surfaces, or through pairs in the bulk (Fig.~\ref{fig:overview}) \cite{henderson2023three,Birch2021Topological,milde2013unwinding,kagawa2017current}. Recent studies involving transitions through fractional topological objects \cite{yu2018transformation,wang2021meron,leonov2021field,yoshimochi2024multistep,hermosa2023bloch,del2024fractional,wang2021meron,rybakov2025topological} highlight the emergence of additional complex transition pathways governed by hidden three-dimensional couplings in real and emergent space.

  \begin{figure*}[t]
\includegraphics[width =\textwidth]{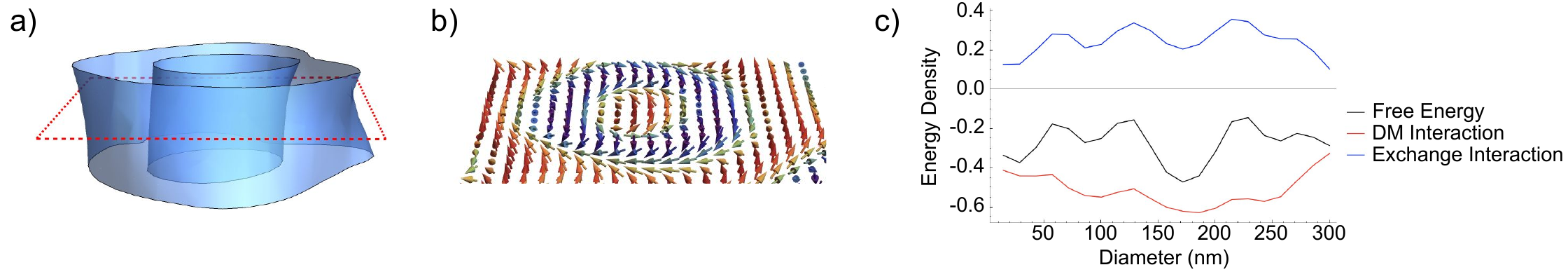}
\caption{ Experimentally observed skyrmionium structure in three dimensions. $M_{z}=-0.5$ contours (a) highlight the doughnut-like spin texture, displayed over a height of 72 nm. Three-dimensional skyrmionium magnetization is shown for one selected depth (b) corresponding to the highlighted region in (a), with the corresponding evolution of the free energy density terms plotted across the width of the skyrmionium (c). The topological charge is calculated from slice b) as 0.006. The width of the skyrmionium is approximately twice the width of a regular skyrmion for this system. Symmetric double-peak structures characterize the transition between opposite in-plane spin rotations when traversing the boundary between inner and outer skyrmions with opposite topological charge.      
}
\label{fig:skyrmionium2d}
\end{figure*}

While research efforts have historically been devoted to two-dimensional skyrmion objects of integer charge, recent advances in three-dimensional materials engineering and imaging techniques have sparked interest in the formation and dynamics of higher-dimensional topological configurations in pursuit of realizing three-dimensional spintronic devices. Here, the interplay of dimensionality and topology couples strings and monopoles in three dimensions to generate novel states with unique functional properties \cite{zheng2023hopfion,Zheng2018experimental, henderson2023three,grelier2022three}. For example, skyrmioniums \cite{zheng2017direct,finazzi2013laser} and hopfions \cite{zheng2023hopfion,kent2021creation} naturally support the unidirectional and ultra-fast transport of topological bits along parallel channels \cite{tang2021magnetic,zhang2016control,zheng2023hopfion,muller2017magnetic,xing2020magnetic,leonov2023swirling}, while skyrmions and hybridized structures  \cite{henderson2023three,Zheng2018experimental} provide novel opportunities for helical controls and multi-bit depth-dependent encoding schemes \cite{Zheng2018experimental,redies2019distinct,henderson2024quantum,chen2023encoding}. Unfortunately, experimental observations of three-dimensional topological structures and their transition pathways remain largely elusive, owing to a lack of imaging techniques suitable for unrestricted sample geometries dominated by bulk energetics. 

Traditional thin-film microscopy techniques yield planar projections whose two-dimensional contrast ambiguity prohibits the direct identification of three-dimensional structures \cite{yang2023reversible,zhang2018real,tang2021magnetic,zheng2023hopfion}, while modern tomography approaches utilizing X-rays and electrons are limited to geometry-constrained samples dominated by surface effects \cite{wolf2022unveiling,seki2022direct,raftrey2024quantifying,yasin2024bloch}. True uninhibited three-dimensional access to bulk skyrmion systems was only recently achieved via a small angle neutron scattering tomography technique presented in \cite{henderson2023three,heacock2020neutron}. Application of this technique to a triangular lattice thermal equilibrium phase in Co$_8$Zn$_8$Mn$_4$ provided unprecedented access to defect-mediated skyrmion topological transition pathways and metastable dipole structures \cite{henderson2023three}. Here, we apply this technique across various field-induced skyrmion dynamics to uncover a zoology of exotic three-dimensional metastable topological structures and novel transition dynamics. Meron and skyrmion bundles are observed throughout the reconstructions, forming $Q = 0$ bound meron and skyrmionium states (Fig.~\ref{fig:overview}), unprecedented in bulk systems.  
Interwoven vortex-antivortex lattices are found to be embedded throughout the entire skyrmion phase, facilitating a new topological transition pathway driven by the coupling of merons and monopoles (Fig.~\ref{fig:overview}). Together, these results unveil novel topological structures and their three-dimensional transition dynamics which open the door to extended information encoding and control schemes across tunable spatial dimensions and arbitrary topological charge for next-generation spintronic devices.

\section{\label{sec:level2}Results\protect\\}

Tomographic multi-projection small angle neutron scattering (SANS) measurements were performed on GP-SANS at the High Flux Isotope Reactor at Oak Ridge National Laboratory \cite{wignall201240,heller2018suite} for a wavelength of 8 \AA. Measurements were collected at a constant temperature of 310 K within the thermal equilibrium triangular lattice skyrmion phase of a disordered Co$_8$Zn$_8$Mn$_4$ bulk sample. At this temperature, tomography was performed for disordered skyrmion states at 0.015 T and 0.025 T upon field-increasing nucleation trajectories, while ordered states were measured at 0.025 T and 0.035 T upon field-increasing skyrmion annihilation trajectories.  Ordered states were achieved via a field-rotating skyrmion ordering procedure outlined in \cite{gilbert2019precipitating,henderson2021characterization,henderson2022skyrmion}, in which the sample was rotated symmetrically in the external magnetic field to precipitate rotationally oriented and ordered skyrmion lattices. Tomography was performed over two axes (rotation and tilt axes shown in Supplementary Fig.~\ref{supp-fig:schematic}), with both the sample and the magnetic field rotated together.  Supporting Monte Carlo simulations were performed for disordered state reconstructions; Supplementary Fig.~\ref{supp-fig:schematic} outlines a comparison of tomographic reconstruction versus simulation processes.

\begin{figure*}
\includegraphics[width =\textwidth]{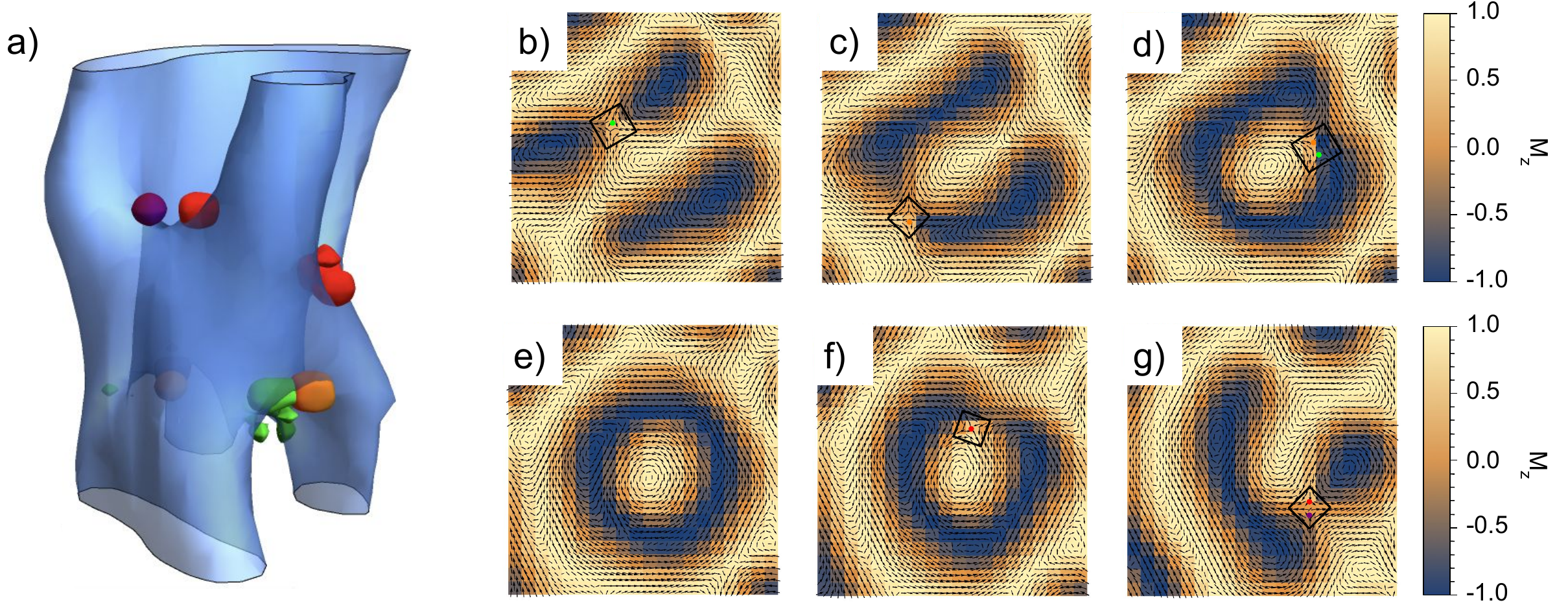}
\caption{Experimentally observed skyrmionium topological transition in three dimensions.  Three-dimensional contours (a) of the z-component of the magnetization, $m_{z} = -0.5$, and the emergent magnetic charge density, $\rho_{em}$, show a coupling of merons and monopoles, with bimerons occurring through the coupling of S$^{-}$ (orange)/B$^{+}$ (green) and S$^{+}$ (purple)/B$^{-}$ (red) (anti)monopoles. In-plane magnetization slices (b-g) show the merging and dissociation of multiple skyrmions through merons and bimerons. Black boxes highlight antivortex and meron topological transition points  with circular dots indicating monopole/antimonopole locations, colored according to their branching/segmenting genre}.  
\label{fig:skyrmionium_3d}
\end{figure*}

Disordered skyrmion reconstructions at 0.025 T, which defines the field for maximal skyrmion formation, exhibit a variety of exotic structures. These structures span $Q = 0$ skyrmioniums (Fig.~\ref{fig:skyrmionium2d}) to bound meron-antimeron pairs (Supplementary Fig.~\ref{supp-fig:meron_antimeron})---structures which have not yet been observed in three dimensions, nor simulated for bulk systems. Skyrmioniums can be described as a composite pair formed by the enclosing of $Q = 1$ and $Q = -1$ skyrmions of opposite magnetization, i.e., opposite polarity \cite{zhang2016control}. In limited two-dimensional imaging geometries, skyrmioniums are typically identified through ring-like regions of magnetic contrast and reversed magnetization \cite{yang2023reversible,zhang2018real,tang2021magnetic}. Unfortunately, this two-dimensional manifestation is not unique to skyrmioniums, with hopfions exhibiting similar features in two dimensions. In our experimental reconstructions (Fig.~\ref{fig:skyrmionium2d}), however, skyrmioniums manifest in three dimensions as double-core cylindrical $m_{z} = -0.5 $ contours (Fig.~\ref{fig:skyrmionium2d}a), where inner and outer contours represent opposite topological charge skyrmions separated by a distance equal to the characteristic skyrmion diameter for the system---approximately 140 nm for the thermal equilibrium triangular lattice phase in Co$_8$Zn$_8$Mn$_4$. In two dimensions (Fig.~\ref{fig:skyrmionium2d}b), these same skyrmioniums exhibit more traditional planar doughnut-like structures in the magnetization slices, reminiscent of previous planar observations made using microscopy techniques \cite{yang2023reversible,zhang2018real}. Tracing the spin orientation from the core to the periphery in two dimensions, we observe a 2$\pi$ vortex where the winding direction of the $Q = 1$ skyrmion reverses at some radii, flipping its polarity to yield a $Q = -1$ skyrmion. The total net topological charge over the skyrmionium's entire diameter therefore equals 0.

 Free energy slices along the horizontal diameter of the skyrmionium, depicted in (Fig.~\ref{fig:skyrmionium2d}c), show the entire width of the skyrmionium to be roughly two times that of a typical skyrmion for the system. Symmetric double-peak structures characterize the transition between opposite in-plane spin rotations when traversing the boundary between inner and outer skyrmions with opposite topological charge. Local minima coincide with locations where the skyrmion polarity is reversed, while a global minimum is observed only at the core of the skyrmionium.

 The evolution of the skyrmionium as a function of depth is experimentally visualized for the first time as depicted in (Fig.~\ref{fig:skyrmionium_3d}). Three-dimensional contours of the z-component of the magnetization, $m_{z}$, and the emergent magnetic charge density, $\rho_{em}$, are shown in a), with segmenting and branching MPs and AMPs colored as S$^{+}$ (purple), S$^{-}$ (orange), B$^{+}$ (green), and B$^{-}$ (red). Corresponding two-dimensional magnetization slices are shown at selected depths along a), where slices b-g) translate vertically from the bottom to the top of the skyrmionium structure. The skyrmion-skyrmionium transition is revealed to occur via the merging of 4 separate skyrmions; two sets of skyrmions first merge to form a pair of elongated skyrmions, which subsequently merge into the concentric circular shape assumed by a skyrmionium. Black boxes highlight anti-vortex locations that develop polarity at merging and separation points, transforming into meron bundles with opposite vorticity, so-called bimerons \cite{yu2024spontaneous}. These bimeron locations correspond to coupled MP-AMP locations in a).

 \begin{figure*}
\includegraphics[width =130mm]{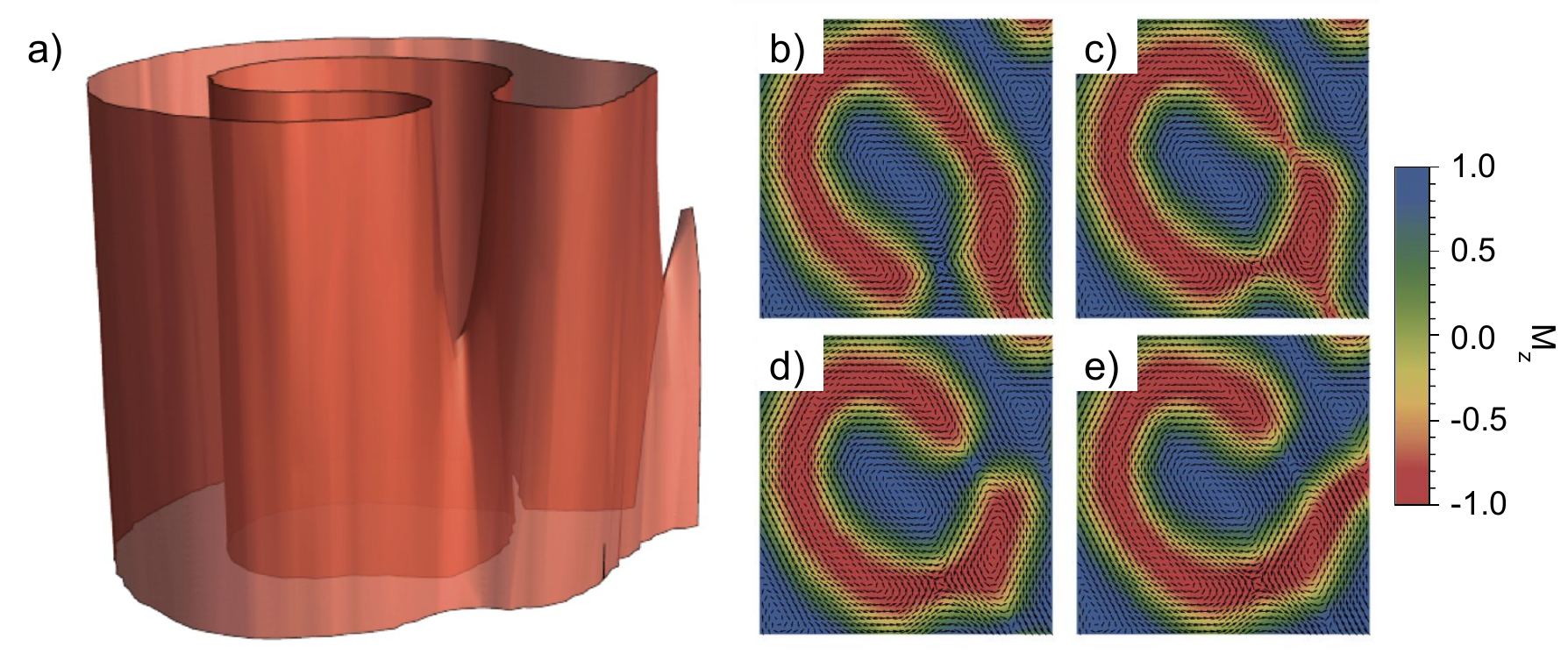}
\caption{Characteristic three-dimensional skyrmionium transition from Monte Carlo simulations. Three-dimensional contours (a) of the z-component of the magnetization, $m_{z} = -0.5$. In-plane magnetization slices (b-g) show similar transition dynamics to (Fig.~\ref{fig:skyrmionium_3d}) via  vortices and merons at transition points.          
}
\label{fig:skyrmionium_sim}
\end{figure*}

 Monte Carlo simulations of the disordered skyrmion state revealed skyrmionium structures with similar transitions occurring through vortices with fractional topological charges (Fig.~\ref{fig:skyrmionium_sim}). Here, a randomly varying exchange field was incorporated to mimic the site disorder present in the sample, in addition to other skyrmion stabilizing interactions such as bulk-type DMI, cubic anisotropy, and Zeeman terms. Field-induced perturbations, through in-plane ordering processes, were shown to eliminate the structures, suggesting the existence of the skyrmion bundles to be metastable in nature, consistent with experimental observations.

\begin{figure}
\includegraphics[width =\columnwidth]{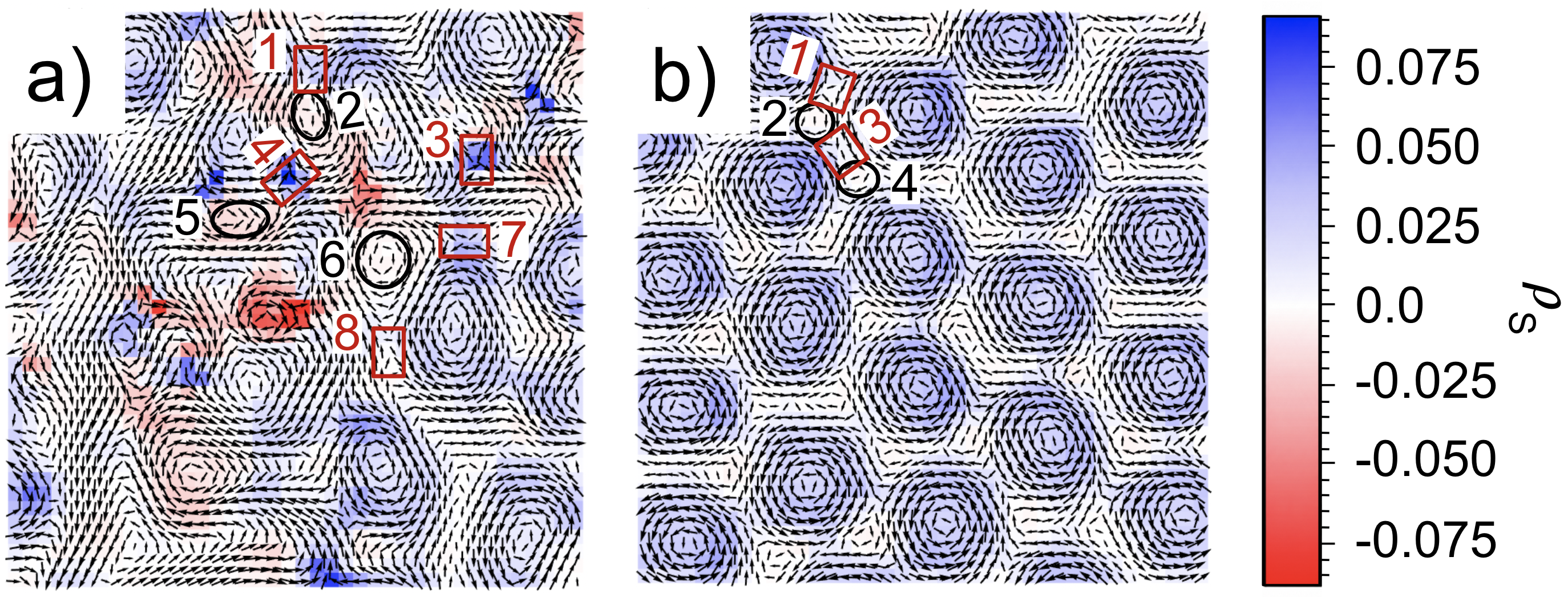}
\caption{ Coexisting skyrmion and vortex-antivortex lattice structures for field-increasing experimental reconstructions. In-plane slices show the topological charge density for selected skyrmion hosting fields, corresponding to the disordered skyrmion state at 0.015 T (a) and the ordered skyrmion state at 0.025 T (b). Black arrows represent the xy component of spin and $\rho_{s}$ denotes the calculated topological charge density. Selected regions containing vortices (black) and antivortices (red) are highlighted, with corresponding topological charges calculated in Table \ref{tab:table}.          
}
\label{fig:vortex}
\end{figure}

To elucidate the origin and role of the vortex-antivortex lattices, we examine field and disorder dependent reconstructions. Fig.~\ref{fig:vortex} outlines selected two-dimensional magnetization slices of the field-increasing reconstructions for the disordered 0.015 T skyrmion state (a) and ordered 0.025 T skyrmion state (b). Universal to these reconstructions is the presence of vortex-antivortex structures within the skyrmion phase envelope, irrespective of field or disorder (see Supplementary Fig.~\ref{supp-fig:vortex} for additional field-dependent reconstructions). More notably, ordered states with higher topological charge saturation (Fig.~\ref{fig:vortex}b)---those which more closely resemble ideal triangular skyrmion lattice arrangements---tend to manifest vortex-antivortex structures with vanishing topological charge. Conversely, disordered reconstructions with lower total topological charge (Fig.~\ref{fig:vortex}a) tend to manifest vortices which possess some fractional topological charge. These observations suggest the presence of vortices within skyrmion lattice is not simply benign, but rather actively influences the topological dynamics of skyrmion systems, exchanging their topology with skyrmions to facilitate field-induced transitions through merons and monopoles. Ultimately, the vortex cores provide regions of elevated exchange energy and strong magnetization curvature/spin distortion which effectively lower the nucleation energy barriers of singular three-dimensional defects, such as Bloch points, monopoles, and merons. In this sense, the vortices may be viewed as the geometric precursors and channels through which topology is exchanged, while the coupling of merons and monopoles enforces global topological conservation constraints.

\begin{table}[H]
\centering
\begin{tabular}{|c|c|c|}
\hline
\multicolumn{1}{|c|}{} 
& \multicolumn{1}{c|}{\makecell{\textbf{a} \\ (0.015 T, D)}} 
& \multicolumn{1}{c|}{\makecell{\textbf{b} \\ (0.025 T, O)}} \\
\hline
1 & 0.26 & 0.04 \\
2 & -0.15 & 0 \\
3 & 0.31 & 0.07 \\
4 & 0.3 & 0.02 \\
5 & -0.22 &  \\
6 & -0.13 &  \\
7 & 0.14 &  \\
8 & 0.05 &  \\
\hline
\end{tabular}
\caption{Topological charge for highlighted regions in plots a) and b) of Fig.~\ref{fig:vortex}. Experimental magnetic field settings are listed for each selected 2D slice taken from the reconstructions, with ordered and disordered states denoted by (O) and (D).}
\label{tab:table}
\end{table}

\section{\label{sec:level2}Discussion\protect\\}

The experimental reconstructions presented here serve as the first three-dimensional visualizations of exotic $Q = 0$ bound skyrmion structures, uncovering novel topological transition pathways through the critical coupling of merons and monopoles in three dimensions. Transitions between $Q = 1$ skyrmions and $Q = 0$ skyrmion bundles are observed to occur via the transformation of preexisting antivortex structures into bimerons bounded in height through coupled MP/AMP pairs at merging/separation points. The coupling of merons and monopoles was first proposed to explain confinement mechanisms in QCD contexts \cite{montero2002monopole}, however, its realization far transcends that of quantum field theory \cite{volovik2000monopoles,xu2024engineering,volovik2020string}. In skyrmion systems, vortices and merons can (1) arise naturally through competing interactions and disorder \cite{yu2018transformation,shao2023emergent,yoshimochi2024multistep,yu2024spontaneous}, or (2) be stabilized externally---most commonly near topological transition boundaries using magnetic field tuning. Given their topological nature, merons may---both individually and in pairs---mediate topological transitions in skyrmion systems \cite{yu2024spontaneous}, governed by emergent MPs/AMPs at transition points along the strings direction \cite{tan2024revealing}. This coupling is intrinsically three-dimensional, linking geometry with topology.     

While skyrmionium formation has been proposed through a myriad of mechanisms spanning boundary confinement effects \cite{zheng2017direct}, interfacial engineering of system parameters \cite{zhang2018real,xia2020current}, and external perturbations \cite{finazzi2013laser,yang2023reversible,zhang2016control}, the systems of interest were primarily limited to ferromagnetic nanostructures where sample depths are traditionally on the order of the skyrmion lattice periodicity. 
Consequently, skyrmionium formation in three-dimensional cubic chiral magnets has remained largely unexplored. Existing studies of skyrmion systems hosting bulk-type DMI's have been shown to be capable of naturally supporting skyrmionium structures \cite{rybakov2019chiral,zheng2022skyrmion}. More recently, the introduction of random spin disorder was also proposed as a possible mechanism for generating skyrmioniums in systems with bulk DMI \cite{zhang2024spin}. We suggest the pervasive nature of vortices within the skyrmion phase and notable presence of $Q = 0$ skyrmion structures, such as skyrmioniums, may be rooted in a combination of nucleation processes and internal disorder within the system. 

Co$_8$Zn$_8$Mn$_4$ constitutes a bulk cubic chiral magnet where a similar spin disorder exists due to site-mixing between ferromagnetic Co spins and antiferromagnetic Mn spins \cite{Karube2020Metastable}. This spin disorder introduces random local variations to the system's exchange parameters, which may stabilize jammed chiral labyrinth domains \cite{henderson2022skyrmion}. Here, the influence of defects and pinning may introduce lower energy barriers for disordered domains, such as ring-like domains, which subsequently serve as nucleation points for exotic $Q = 0$ skyrmion structures upon field ramping into the skyrmion window. Moreover, the unique observation of exotic $Q = 0$ skyrmion bundles within the disordered 0.025 T skyrmion state highlights the metastable nature of these objects, wherein annihilation occurs upon perturbations in field direction and magnitude. Application of the ordering sequence likely disrupts the states, forcing their decay into $Q = 1$ skyrmions and vortices which serve as the remnants of the interior $Q = -1$ skyrmion structures. Similar instabilities and destruction pathways have been demonstrated for simulated skyrmioniums in tilted magnetic fields \cite{zheng2022skyrmion}. 
The simulations presented in Fig.~\ref{fig:skyrmionium_sim} reinforce this metastability.

Similar observations of composite skyrmion and vortex structures, reminiscent of those found in our reconstructions, are limited to micromagnetic simulations involving the attraction and coupling of horizontal and vertical skyrmion tubes during skyrmion nucleation processes \cite{leonov2021field,vlasov2020skyrmion,leonov2023swirling}. Horizontal skyrmion tubes---aligned perpendicular to the field---were first identified in simulations of low-field conical phases of cubic helimagnets, constituting an intermediate phase of attractive skyrmions \cite{leonov2021field,vlasov2020skyrmion}. Crossover regions between horizontal and vertical tube configurations have since uncovered complex skyrmion clusters and composite objects, encompassing spin solenoids \cite{leonov2021field} to hopfions \cite{leonov2023swirling}. Notably, the horizontal skyrmions can be decomposed into bimerons, stabilizing opposite $Q$ vertical skyrmions and skyrmioniums through the interactions of horizontal and vertical skyrmions \cite{leonov2023swirling}. More generally, merons represent ruptures in the spiral state whose condensation stabilizes both individual triangular lattices of merons and skyrmions, in addition to intercalated lattices of the two at various fields \cite{mukai2022skyrmion,leonov2021field}. The mutual attraction responsible for this condensation process may arise from interacting horizontal tubes, with the intercalated lattice forming a local energy minimum \cite{mukai2022skyrmion,leonov2021field}. Given the similarity in structures between our reconstructions and these simulations, it is possible that our observed composite vortex and skyrmion structures provide evidence of skyrmion nucleation processes through intermediate states of horizontal tubes within our sample. Here, the presence of exotic $Q = 0$ structures and bimeron-facilitated transitions could be seen as remnant structures from the interactions of horizontal and vertical skyrmion tubes in crossover field regions, while the presence of compact interwoven meron and vortex lattices can be understood as energetically favorable configurations produced as a result of the dissociation of horizontal skyrmions into meron pairs during field-increasing nucleation processes. These merons may also mediate the transition from disordered skyrmion to ordered skyrmion states at 0.025 T, increasing the topological saturation of the system and leaving behind topologically trivial vortices in their place. Such topological dynamics are further supported by high-field reconstructions which highlight trivial vortex structures (Supplementary Fig.~\ref{supp-fig:vortex} e) that may be interpreted as remnants from meron-assisted pathways during the topological annihilation of the skyrmion lattice.

The realization of three-dimensional $Q = 0$ skyrmion bundles and their meron-facilitated dynamics in free geometries opens the door to a new genre of three-dimensional physics for spintronic device applications \cite{henderson2024quantum}. For example, the topological objects presented here spanning $Q = 0$ skyrmioniums, $Q=\pm \frac{1}{2}$ merons, and $Q=\pm 1$ skyrmions may be exploited in future frameworks to develop multi-bit encoding schemes through sequences of topological bits. Furthermore, controlled parallel transport may be achieved in these systems through individual skyrmionium motion or via meron propagation in confined channels, such as helical backgrounds or horizontal skyrmion tubes. Finally, disorder has been shown to act as a tunable control parameter which stabilizes diverse topological structures and drives meron and monopole dynamics in three dimensions. This collection of three-dimensional phenomena inspires novel topological encoding and manipulation schemes which realize a new generation of spintronic devices, free from the traditional constraints that limit existing two-dimensional skyrmion architectures.

\begin{acknowledgments}

This work was supported by the Canadian Excellence Research Chairs (CERC) program, the Natural Sciences and Engineering Council of Canada (NSERC) Discovery program, the Canada First Research Excellence Fund (CFREF), and the U.S. Department of Energy, Office of Nuclear Physics, under Interagency Agreement 89243019SSC000025. Part of the work was supported by the U.S. Department of Energy (DOE), Office of Science, Office of Basic Energy Sciences, Early Career Research Program Award KC0402010, under Contract DE-AC05-00OR22725. A portion of this research used resources at the High Flux Isotope Reactor, a DOE Office of Science User Facility operated by the Oak Ridge National Laboratory. The beamtime was allocated to GP-SANS on proposal number IPTS-26872.1. This work is based partly on experiments performed at the Swiss spallation neutron source SINQ, Paul Scherrer Institute, Villigen, Switzerland. O.A.T. acknowledges support from the Australian Research Council (Grant No. DP240101062), NCMAS grant, and visiting program of ICC-IMR, Tohoku University (Japan). 

\end{acknowledgments}

\bibliography{references}

@ARTICLE{milde2013unwinding,
  title={Unwinding of a skyrmion lattice by magnetic monopoles},
  author={Milde, Peter and K{\"o}hler, Denny and Seidel, Joachim and Eng, LM and Bauer, Andreas and Chacon, Alfonso and Kindervater, Jonas and M{\"u}hlbauer, Sebastian and Pfleiderer, Christian and Buhrandt, Stefan and others},
  journal={Science},
  volume={340},
  number={6136},
  pages={1076--1080},
  year={2013},
  publisher={American Association for the Advancement of Science}
}

@ARTICLE{kagawa2017current,
  title={Current-induced viscoelastic topological unwinding of metastable skyrmion strings},
  author={Kagawa, Fumitaka and Oike, Hiroshi and Koshibae, Wataru and Kikkawa, Akiko and Okamura, Yoshihiro and Taguchi, Yasujiro and Nagaosa, Naoto and Tokura, Yoshinori},
  journal={Nature Communications},
  volume={8},
  number={1},
  pages={1332},
  year={2017},
  publisher={Nature Publishing Group}
}

@ARTICLE{gilbert2019precipitating,
title={Precipitating ordered skyrmion lattices from helical spaghetti and granular powders},
  author={Gilbert, Dustin A and Grutter, Alexander J and Neves, Paul M and Shu, Guo-Jiun and Zimanyi, Gergely and Maranville, Brian B and Chou, Fang-Cheng and Krycka, Kathryn and Butch, Nicholas P and Huang, Sunxiang and others},
  journal={Physical Review Materials},
  volume={3},
  number={1},
  pages={014408},
  year={2019},
  publisher={APS}
}

@article{heacock2020neutron,
  title={Neutron sub-micrometre tomography from scattering data},
  author={Heacock, B and Sarenac, D and Cory, DG and Huber, MG and MacLean, JPW and Miao, H and Wen, H and Pushin, DA},
  journal={IUCrJ},
  volume={7},
  number={5},
  pages={893--900},
  year={2020},
  publisher={International Union of Crystallography}
}

@article{yu2018transformation,
  title={Transformation between meron and skyrmion topological spin textures in a chiral magnet},
  author={Yu, XZ and Koshibae, W and Tokunaga, Y and Shibata, K and Taguchi, Y and Nagaosa, N and Tokura, Y},
  journal={Nature},
  volume={564},
  number={7734},
  pages={95--98},
  year={2018},
  publisher={Nature Publishing Group}
}

@article{schutte2014dynamics,
  title={Dynamics and energetics of emergent magnetic monopoles in chiral magnets},
  author={Schütte, Christoph and Rosch, Achim},
  journal={Physical Review B},
  volume={90},
  number={17},
  pages={174432},
  year={2014},
  publisher={APS}
}

@article{Zhang2015magnetic,
  title={Magnetic skyrmion logic gates: conversion, duplication and merging of skyrmions},
  author={Zhang, Xichao and Ezawa, Motohiko and Zhou, Yan},
  journal={Scientific Reports},
  volume={5},
  number={1},
  pages={9400},
  year={2015},
  publisher={Nature Publishing Group}
}

@article{Fert2017magnetic,
  title={Magnetic skyrmions: advances in physics and potential applications},
  author={Fert, Albert and Reyren, Nicolas and Cros, Vincent},
  journal={Nature Review Materials},
  volume={2},
  number={7},
  pages={17031},
  year={2017},
  publisher={Nature Publishing Group}
}

@article{Schulz2012emergent,
  title={Emergent electrodynamics of skyrmions in a chiral magnet},
  author={Schulz, T. and Ritz, R. and Bauer, A. and Halder, M. and Wagner, M. and Franz, C. and Pfleiderer, C. and Everschor, K. and Garst, M. and Rosch, A.},
  journal={Nature Physics},
  volume={8},
  number={4},
  pages={301--304},
  year={2012},
  publisher={IOP Publishingg}
}

@article{nagaosa2012gauge,
  title={Gauge fields in real and momentum spaces in magnets: monopoles and skyrmions},
  author={Nagaosa, N and Yu, XZ and Tokura, Y},
  journal={Philosophical Transactions of the Royal Society A: Mathematical, Physical and Engineering Sciences},
  volume={370},
  number={1981},
  pages={5806--5819},
  year={2012},
  publisher={The Royal Society Publishing}
}

@article{everschor2014real,
  title={Real-space Berry phases: Skyrmion soccer},
  author={Everschor-Sitte, Karin and Sitte, Matthias},
  journal={Journal of Applied Physics},
  volume={115},
  number={17},
  year={2014},
  publisher={AIP Publishing}
}

@article{Yu2010real,
  title={Real-space observation of a two-dimensional skyrmion crystal},
  author={Yu, X. Z. and Onose, Y. and Kanazawa, N. and Park, J. H. and Han, J. H. and Matsui, Y. and Nagaosa, N. and Tokura, Y.},
  journal={Nature},
  volume={465},
  number={7300},
  pages={901--904},
  year={2010},
  publisher={Nature Publishing Group}
}

@article{seki2022direct,
  title={Direct visualization of the three-dimensional shape of skyrmion strings in a noncentrosymmetric magnet},
  author={Seki, S and Suzuki, M and Ishibashi, M and Takagi, R and Khanh, ND and Shiota, Y and Shibata, K and Koshibae, W and Tokura, Y and Ono, T},
  journal={Nature Materials},
  volume={21},
  number={2},
  pages={181--187},
  year={2022},
  publisher={Nature Publishing Group UK London}
}

@article{Yu2020real,
  title = {Real-Space Observation of Topological Defects in Extended Skyrmion-Strings},
  author = {Yu, X. and Masell, J. and Yasin, F. S. and Karube, K. and Kanazawa, N. and Nakajima, K. and Nagai, T. and Kimoto, K. and Koshibae, W. and Taguchi, Y. and Nagaosa, N. and Tokura, Y. },
  journal = {Nano Letters},
  volume = {20},
  issue = {10},
  pages = {7313-7320},
  year = {2020},
  doi = {10.1021/acs.nanolett.0c02708}
}

@article{Zheng2018experimental,
  title = {Experimental observation of chiral magnetic bobbers in {B20}-type {FeGe}},
  author = {Zheng, F. and Rybakov, F. N. and Borisov, A. B. and Song, D. and Wang, S. and Li, Z. A.  and Du, H. and Kiselev, N. S. and Caron, J. and Kovacs, A. and Tian, M. and Zhang, Y. and Blugel, S. and Dunin-Borkowski, R. E. },
  journal = {Nature Nanotechnology},
  volume = {13},
  issue = {6},
  pages = {451-455},
  year = {2018},
  doi = {10.1038/s41565-018-0093-3}
}

@article{jin2023evolution,
  title={Evolution of Emergent Monopoles into Magnetic Skyrmion Strings},
  author={Jin, Haonan and Tan, Wancong and Liu, Yizhou and Ran, Kejing and Fan, Raymond and Shangguan, Yanyan and Guang, Yao and van der Laan, Gerrit and Hesjedal, Thorsten and Wen, Jinsheng and others},
  journal={Nano Letters},
  year={2023},
  publisher={ACS Publications}
}

@article{Birch2021Topological,
author = {Birch, M. T. and Cort{\'e}s-Ortu{\~n}o, D. and Khanh, N. D. and Seki, S. and \ifmmode \check{S}\else \v{S}\fi{}tefan\ifmmode \check{c}\else \v{c}\fi{}i\ifmmode \check{c}\else \v{c}\fi{}, A. and Balakrishnan, G. and Tokura, Y. and Hatton, P. D.},
title = {Topological defect-mediated skyrmion annihilation in three dimensions},
journal = {Communication Physics},
volume = {4},

pages = {175},
year = {2021},
}

@article{Karube2020Metastable,
  title = {Metastable skyrmion lattices governed by magnetic disorder and anisotropy in $\ensuremath{\beta}$-Mn-type chiral magnets},
  author = {Karube, K. and White, J. S. and Ukleev, V. and Dewhurst, C. D. and Cubitt, R. and Kikkawa, A. and Tokunaga, Y. and R\o{}nnow, H. M. and Tokura, Y. and Taguchi, Y.},
  journal = {Phys. Rev. B},
  volume = {102},
  issue = {6},
  pages = {064408},
  numpages = {20},
  year = {2020},
  publisher = {American Physical Society},
  doi = {10.1103/PhysRevB.102.064408},
  
}

@article{wolf2022unveiling,
  title={Unveiling the three-dimensional magnetic texture of skyrmion tubes},
  author={Wolf, Daniel and Schneider, Sebastian and R{\"o}{\ss}ler, Ulrich K and Kov{\'a}cs, Andr{\'a}s and Schmidt, Marcus and Dunin-Borkowski, Rafal E and B{\"u}chner, Bernd and Rellinghaus, Bernd and Lubk, Axel},
  journal={Nature Nanotechnology},
  volume={17},
  number={3},
  pages={250--255},
  year={2022},
  publisher={Nature Publishing Group UK London}
}

@article{henderson2023three,
  title={Three-dimensional neutron far-field tomography of a bulk skyrmion lattice},
  author={Henderson, ME and Heacock, B and Bleuel, M and Cory, DG and Heikes, C and Huber, MG and Krzywon, J and Nahman-Levesqu{\'e}, O and Luke, GM and Pula, M and others},
  journal={Nature Physics},
  pages={1--7},
  year={2023},
  publisher={Nature Publishing Group UK London}
}

@article{henderson2022skyrmion,
  title={Skyrmion alignment and pinning effects in the disordered multiphase skyrmion material Co 8 Zn 8 Mn 4},
  author={Henderson, ME and Bleuel, M and Beare, J and Cory, DG and Heacock, B and Huber, MG and Luke, GM and Pula, M and Sarenac, D and Sharma, S and others},
  journal={Physical Review B},
  volume={106},
  number={9},
  pages={094435},
  year={2022},
  publisher={APS}
}

@article{henderson2021characterization,
  title={Characterization of a Disordered above Room Temperature Skyrmion Material {$\mathrm{Co}_{8}\mathrm{Zn}_{8}\mathrm{Mn}_{4}$}},
  author={Henderson, Melissa E and Beare, James and Sharma, Sudarshan and Bleuel, Markus and Clancy, Pat and Cory, David G and Huber, Michael G and Marjerrison, Casey A and Pula, Mathew and Sarenac, Dusan and others},
  journal={Materials},
  volume={14},
  number={16},
  pages={4689},
  year={2021},
  publisher={Multidisciplinary Digital Publishing Institute}
}

@article{wachowiak2002direct,
  title={Direct observation of internal spin structure of magnetic vortex cores},
  author={Wachowiak, A and Wiebe, J and Bode, M and Pietzsch, O and Morgenstern, M and Wiesendanger, R},
  journal={Science},
  volume={298},
  number={5593},
  pages={577--580},
  year={2002},
  publisher={American Association for the Advancement of Science}
}

@article{yasin2024bloch,
  title={Bloch point quadrupole constituting hybrid topological strings revealed with electron holographic vector field tomography},
  author={Yasin, Fehmi Sami and Masell, Jan and Takahashi, Yoshio and Akashi, Tetsuya and Baba, Norio and Karube, Kosuke and Shindo, Daisuke and Arima, Takahisa and Taguchi, Yasujiro and Tokura, Yoshinori and others},
  journal={Advanced Materials},
  volume={36},
  number={16},
  pages={2311737},
  year={2024},
  publisher={Wiley Online Library}
}

@article{zheng2023hopfion,
  title={Hopfion rings in a cubic chiral magnet},
  author={Zheng, Fengshan and Kiselev, Nikolai S and Rybakov, Filipp N and Yang, Luyan and Shi, Wen and Bl{\"u}gel, Stefan and Dunin-Borkowski, Rafal E},
  journal={Nature},
  volume={623},
  number={7988},
  pages={718--723},
  year={2023},
  publisher={Nature Publishing Group UK London}
}

@article{grelier2022three,
  title={Three-dimensional skyrmionic cocoons in magnetic multilayers},
  author={Grelier, Matthieu and Godel, Florian and Vecchiola, Aymeric and Collin, Sophie and Bouzehouane, Karim and Fert, Albert and Cros, Vincent and Reyren, Nicolas},
  journal={Nature Communications},
  volume={13},
  number={1},
  pages={6843},
  year={2022},
  publisher={Nature Publishing Group UK London}
}

@article{hermosa2023bloch,
  title={Bloch points and topological dipoles observed by X-ray vector magnetic tomography in a ferromagnetic microstructure},
  author={Hermosa, Javier and Hierro-Rodr{\'\i}guez, Aurelio and Quir{\'o}s, Carlos and Mart{\'\i}n, Jos{\'e} I and Sorrentino, Andrea and Aballe, Luc{\'\i}a and Pereiro, Eva and V{\'e}lez, Mar{\'\i}a and Ferrer, Salvador},
  journal={Communications Physics},
  volume={6},
  number={1},
  pages={49},
  year={2023},
  publisher={Nature Publishing Group UK London}
}

@article{wang2021meron,
  title={Meron, skyrmion, and vortex crystals in centrosymmetric tetragonal magnets},
  author={Wang, Zhentao and Su, Ying and Lin, Shi-Zeng and Batista, Cristian D},
  journal={Physical Review B},
  volume={103},
  number={10},
  pages={104408},
  year={2021},
  publisher={APS}
}

@article{gao2019creation,
  title={Creation and annihilation of topological meron pairs in in-plane magnetized films},
  author={Gao, Ningbo and Je, S-G and Im, M-Y and Choi, Jun Woo and Yang, Masheng and Li, Qin-ci and Wang, TY and Lee, S and Han, H-S and Lee, K-S and others},
  journal={Nature Communications},
  volume={10},
  number={1},
  pages={5603},
  year={2019},
  publisher={Nature Publishing Group UK London}
}

@article{montero2002monopole,
  title={Monopole and vortex content of a meron pair},
  author={Montero, Alvaro and Negele, John W},
  journal={Physics Letters B},
  volume={533},
  number={3-4},
  pages={322--329},
  year={2002},
  publisher={Elsevier}
}

@article{zhang2024spin,
  title={Spin disorder control of topological spin texture},
  author={Zhang, Hongrui and Shao, Yu-Tsun and Chen, Xiang and Zhang, Binhua and Wang, Tianye and Meng, Fanhao and Xu, Kun and Meisenheimer, Peter and Chen, Xianzhe and Huang, Xiaoxi and others},
  journal={Nature Communications},
  volume={15},
  number={1},
  pages={3828},
  year={2024},
  publisher={Nature Publishing Group UK London}
}

@article{al2001skyrmions,
  title={Skyrmions in a ferromagnetic Bose--Einstein condensate},
  author={Al Khawaja, Usama and Stoof, Henk},
  journal={Nature},
  volume={411},
  number={6840},
  pages={918--920},
  year={2001},
  publisher={Nature Publishing Group UK London}
}

@article{kleman2006topological,
  title={Topological point defects in nematic liquid crystals},
  author={Kleman, Maurice and Lavrentovich, Oleg D},
  journal={Philosophical Magazine},
  volume={86},
  number={25-26},
  pages={4117--4137},
  year={2006},
  publisher={Taylor \& Francis}
}

@article{chuang1991cosmology,
  title={Cosmology in the laboratory: Defect dynamics in liquid crystals},
  author={Chuang, Isaac and Durrer, Ruth and Turok, Neil and Yurke, Bernard},
  journal={Science},
  volume={251},
  number={4999},
  pages={1336--1342},
  year={1991},
  publisher={American Association for the Advancement of Science}
}

@article{kleman2008disclinations,
  title={Disclinations, dislocations, and continuous defects: A reappraisal},
  author={Kleman, Maurice and Friedel, Jacques},
  journal={Reviews of Modern Physics},
  volume={80},
  number={1},
  pages={61--115},
  year={2008},
  publisher={APS}
}

@article{zhang2016control,
  title={Control and manipulation of a magnetic skyrmionium in nanostructures},
  author={Zhang, Xichao and Xia, Jing and Zhou, Yan and Wang, Daowei and Liu, Xiaoxi and Zhao, Weisheng and Ezawa, Motohiko},
  journal={Physical Review B},
  volume={94},
  number={9},
  pages={094420},
  year={2016},
  publisher={APS}
}

@article{muller2017magnetic,
  title={Magnetic skyrmions and skyrmion clusters in the helical phase of cu 2 oseo 3},
  author={M{\"u}ller, Jan and Rajeswari, Jayaraman and Huang, Ping and Murooka, Yoshie and R{\o}nnow, Henrik M and Carbone, Fabrizio and Rosch, Achim},
  journal={Physical Review Letters},
  volume={119},
  number={13},
  pages={137201},
  year={2017},
  publisher={APS}
}

@article{henderson2024quantum,
  title={Quantum Advancements in Neutron Scattering Reshape Spintronic Devices},
  author={Henderson, ME and Cory, DG and Sarenac, D and Pushin, DA},
  journal={arXiv preprint arXiv:2407.10822},
  year={2024}
}

@article{chen2023encoding,
  title={Encoding and multiplexing information signals in magnetic multilayers with fractional skyrmion tubes},
  author={Chen, Runze and Li, Yu and Griggs, Will and Zang, Yuzhe and Pavlidis, Vasilis F and Moutafis, Christoforos},
  journal={ACS Applied Materials \& Interfaces},
  volume={15},
  number={28},
  pages={34145--34158},
  year={2023},
  publisher={ACS Publications}
}

@article{shao2023emergent,
  title={Emergent chirality in a polar meron to skyrmion phase transition},
  author={Shao, Yu-Tsun and Das, Sujit and Hong, Zijian and Xu, Ruijuan and Chandrika, Swathi and G{\'o}mez-Ortiz, Fernando and Garc{\'\i}a-Fern{\'a}ndez, Pablo and Chen, Long-Qing and Hwang, Harold Y and Junquera, Javier and others},
  journal={Nature Communications},
  volume={14},
  number={1},
  pages={1355},
  year={2023},
  publisher={Nature Publishing Group UK London}
}

@article{yoshimochi2024multistep,
  title={Multistep topological transitions among meron and skyrmion crystals in a centrosymmetric magnet},
  author={Yoshimochi, H and Takagi, R and Ju, J and Khanh, ND and Saito, H and Sagayama, H and Nakao, H and Itoh, S and Tokura, Y and Arima, T and others},
  journal={Nature Physics},
  volume={20},
  number={6},
  pages={1001--1008},
  year={2024},
  publisher={Nature Publishing Group UK London}
}

@article{yu2024spontaneous,
  title={Spontaneous Vortex-Antivortex Pairs and Their Topological Transitions in a Chiral-Lattice Magnet},
  author={Yu, Xiuzhen and Kanazawa, Naoya and Zhang, Xichao and Takahashi, Yoshio and Iakoubovskii, Konstantin V and Nakajima, Kiyomi and Tanigaki, Toshiaki and Mochizuki, Masahito and Tokura, Yoshinori},
  journal={Advanced Materials},
  volume={36},
  number={1},
  pages={2306441},
  year={2024},
  publisher={Wiley Online Library}
}

@article{redies2019distinct,
  title={Distinct magnetotransport and orbital fingerprints of chiral bobbers},
  author={Redies, M and Lux, FR and Hanke, J-P and Buhl, PM and M{\"u}ller, GP and Kiselev, NS and Bl{\"u}gel, S and Mokrousov, Y},
  journal={Physical Review B},
  volume={99},
  number={14},
  pages={140407},
  year={2019},
  publisher={APS}
}

@article{tang2021magnetic,
  title={Magnetic skyrmion bundles and their current-driven dynamics},
  author={Tang, Jin and Wu, Yaodong and Wang, Weiwei and Kong, Lingyao and Lv, Boyao and Wei, Wensen and Zang, Jiadong and Tian, Mingliang and Du, Haifeng},
  journal={Nature Nanotechnology},
  volume={16},
  number={10},
  pages={1086--1091},
  year={2021},
  publisher={Nature Publishing Group UK London}
}

@article{zheng2017direct,
  title={Direct imaging of a zero-field target skyrmion and its polarity switch in a chiral magnetic nanodisk},
  author={Zheng, Fengshan and Li, Hang and Wang, Shasha and Song, Dongsheng and Jin, Chiming and Wei, Wenshen and Kov{\'a}cs, Andr{\'a}s and Zang, Jiadong and Tian, Mingliang and Zhang, Yuheng and others},
  journal={Physical Review Letters},
  volume={119},
  number={19},
  pages={197205},
  year={2017},
  publisher={APS}
}

@article{finazzi2013laser,
  title={Laser-induced magnetic nanostructures with tunable topological properties},
  author={Finazzi, Marco and Savoini, Matteo and Khorsand, AR and Tsukamoto, A and Itoh, A and Duo, Lamberto and Kirilyuk, Andrei and Rasing, Th and Ezawa, M},
  journal={Physical Review Letters},
  volume={110},
  number={17},
  pages={177205},
  year={2013},
  publisher={APS}
}

@article{kent2021creation,
  title={Creation and observation of Hopfions in magnetic multilayer systems},
  author={Kent, Noah and Reynolds, Neal and Raftrey, David and Campbell, Ian TG and Virasawmy, Selven and Dhuey, Scott and Chopdekar, Rajesh V and Hierro-Rodriguez, Aurelio and Sorrentino, Andrea and Pereiro, Eva and others},
  journal={Nature Communications},
  volume={12},
  number={1},
  pages={1562},
  year={2021},
  publisher={Nature Publishing Group UK London}
}

@article{xing2020magnetic,
  title={Magnetic skyrmion tubes as nonplanar magnonic waveguides},
  author={Xing, Xiangjun and Zhou, Yan and Braun, HB},
  journal={Physical Review Applied},
  volume={13},
  number={3},
  pages={034051},
  year={2020},
  publisher={APS}
}

@article{vlasov2020skyrmion,
  title={Skyrmion flop transition and congregation of mutually orthogonal skyrmions in cubic helimagnets},
  author={Vlasov, Sergei M and Uzdin, Valery M and Leonov, Andrey O},
  journal={Journal of Physics: Condensed Matter},
  volume={32},
  number={18},
  pages={185801},
  year={2020},
  publisher={IOP Publishing}
}

@article{leonov2021field,
  title={Field-driven metamorphoses of isolated skyrmions within the conical state of cubic helimagnets},
  author={Leonov, Andrey O and Pappas, Catherine and Smalyukh, Ivan I},
  journal={Physical Review B},
  volume={104},
  number={6},
  pages={064432},
  year={2021},
  publisher={APS}
}

@article{leonov2023swirling,
  title={Swirling of Horizontal Skyrmions into Hopfions in Bulk Cubic Helimagnets},
  author={Leonov, Andrey O},
  journal={Magnetism},
  volume={3},
  number={4},
  pages={297--307},
  year={2023},
  publisher={MDPI}
}

@article{tan2024revealing,
  title={Revealing emergent magnetic charge in an antiferromagnet with diamond quantum magnetometry},
  author={Tan, Anthony KC and Jani, Hariom and H{\"o}gen, Michael and Stefan, Lucio and Castelnovo, Claudio and Braund, Daniel and Geim, Alexandra and Mechnich, Annika and Feuer, Matthew SG and Knowles, Helena S and others},
  journal={Nature Materials},
  volume={23},
  number={2},
  pages={205--211},
  year={2024},
  publisher={Nature Publishing Group UK London}
}

@article{zhang2018real,
  title={Real-space observation of skyrmionium in a ferromagnet-magnetic topological insulator heterostructure},
  author={Zhang, Shilei and Kronast, Florian and van der Laan, Gerrit and Hesjedal, Thorsten},
  journal={Nano Letters},
  volume={18},
  number={2},
  pages={1057--1063},
  year={2018},
  publisher={ACS Publications}
}

@article{yang2023reversible,
  title={Reversible conversion between skyrmions and skyrmioniums},
  author={Yang, Sheng and Zhao, Yuelei and Wu, Kai and Chu, Zhiqin and Xu, Xiaohong and Li,f Xiaoguang and {\AA}kerman, Johan and Zhou, Yan},
  journal={Nature Communications},
  volume={14},
  number={1},
  pages={3406},
  year={2023},
  publisher={Nature Publishing Group UK London}
}

@article{xia2020current,
  title={Current-driven skyrmionium in a frustrated magnetic system},
  author={Xia, Jing and Zhang, Xichao and Ezawa, Motohiko and Tretiakov, Oleg A and Hou, Zhipeng and Wang, Wenhong and Zhao, Guoping and Liu, Xiaoxi and Diep, Hung T and Zhou, Yan},
  journal={Applied Physics Letters},
  volume={117},
  number={1},
  year={2020},
  publisher={AIP Publishing}
}

@article{rybakov2019chiral,
  title={Chiral magnetic skyrmions with arbitrary topological charge},
  author={Rybakov, Filipp N and Kiselev, Nikolai S},
  journal={Physical Review B},
  volume={99},
  number={6},
  pages={064437},
  year={2019},
  publisher={APS}
}

@article{zheng2022skyrmion,
  title={Skyrmion--antiskyrmion pair creation and annihilation in a cubic chiral magnet},
  author={Zheng, Fengshan and Kiselev, Nikolai S and Yang, Luyan and Kuchkin, Vladyslav M and Rybakov, Filipp N and Bl{\"u}gel, Stefan and Dunin-Borkowski, Rafal E},
  journal={Nature Physics},
  volume={18},
  number={8},
  pages={863--868},
  year={2022},
  publisher={Nature Publishing Group UK London}
}

@article{heller2018suite,
  title={The suite of small-angle neutron scattering instruments at Oak Ridge National Laboratory},
  author={Heller, William T and Cuneo, Matthew and Debeer-Schmitt, Lisa and Do, Changwoo and He, Lilin and Heroux, Luke and Littrell, Kenneth and Pingali, Sai Venkatesh and Qian, Shuo and Stanley, Christopher and others},
  journal={Journal of Applied Crystallography},
  volume={51},
  number={2},
  pages={242--248},
  year={2018},
  publisher={International Union of Crystallography}
}

@article{wignall201240,
  title={The 40 m general purpose small-angle neutron scattering instrument at Oak Ridge National Laboratory},
  author={Wignall, George D and Littrell, Kenneth C and Heller, William T and Melnichenko, Yuri B and Bailey, Kathy M and Lynn, Gary W and Myles, Dean A and Urban, Volker S and Buchanan, Michelle V and Selby, Douglas L and others},
  journal={Journal of Applied Crystallography},
  volume={45},
  number={5},
  pages={990--998},
  year={2012},
  publisher={International Union of Crystallography}
}

@article{raftrey2024quantifying,
  title={Quantifying the topology of magnetic skyrmions in three dimensions},
  author={Raftrey, David and Finizio, Simone and Chopdekar, Rajesh V and Dhuey, Scott and Bayaraa, Temuujin and Ashby, Paul and Raabe, J{\"o}rg and Santos, Tiffany and Griffin, Sin{\'e}ad and Fischer, Peter},
  journal={Science Advances},
  volume={10},
  number={40},
  pages={eadp8615},
  year={2024},
  publisher={American Association for the Advancement of Science}
}

@article{volovik2000monopoles,
  title={Monopoles and fractional vortices in chiral superconductors},
  author={Volovik, GE},
  journal={Proceedings of the National Academy of Sciences},
  volume={97},
  number={6},
  pages={2431--2436},
  year={2000},
  publisher={National Acad Sciences}
}

@article{xu2024engineering,
  title={Engineering and revealing Dirac strings in spinor condensates},
  author={Xu, Gui-Sheng and Jain, Mudit and Zhou, Xiang-Fa and Guo, Guang-Can and Amin, Mustafa A and Pu, Han and Zhou, Zheng-Wei},
  journal={Physical Review Research},
  volume={6},
  number={2},
  pages={023272},
  year={2024},
  publisher={APS}
}

@article{volovik2020string,
  title={String monopoles, string walls, vortex skyrmions, and nexus objects in the polar distorted B phase of He 3},
  author={Volovik, GE and Zhang, Kuang},
  journal={Physical Review Research},
  volume={2},
  number={2},
  pages={023263},
  year={2020},
  publisher={APS}
}

@article{mukai2022skyrmion,
  title={Skyrmion and meron ordering in quasi-two-dimensional chiral magnets},
  author={Mukai, Natsuki and Leonov, Andrey O},
  journal={Physical Review B},
  volume={106},
  number={22},
  pages={224428},
  year={2022},
  publisher={APS}
}

@article{del2024fractional,
  title={Fractional topological charges in two-dimensional magnets},
  author={del Ser, Nina and El Achchi, Imane and Rosch, Achim},
  journal={Physical Review B},
  volume={110},
  number={9},
  pages={094442},
  year={2024},
  publisher={APS}
}

@article{donnelly2021experimental,
  title={Experimental observation of vortex rings in a bulk magnet},
  author={Donnelly, Claire and Metlov, Konstantin L and Scagnoli, Valerio and Guizar-Sicairos, Manuel and Holler, Mirko and Bingham, Nicholas S and Raabe, J{\"o}rg and Heyderman, Laura J and Cooper, Nigel R and Gliga, Sebastian},
  journal={Nature Physics},
  volume={17},
  number={3},
  pages={316--321},
  year={2021},
  publisher={Nature Publishing Group UK London}
}

@article{rybakov2025topological,
  title={Topological invariants of vortices, merons, skyrmions, and their combinations in continuous and discrete systems},
  author={Rybakov, Filipp N and Eriksson, Olle and Kiselev, Nikolai S},
  journal={Physical Review B},
  volume={111},
  number={13},
  pages={134417},
  year={2025},
  publisher={APS}
}

@article{gobel2021beyond,
  title={Beyond skyrmions: Review and perspectives of alternative magnetic quasiparticles},
  author={G{\"o}bel, B{\"o}rge and Mertig, Ingrid and Tretiakov, Oleg A},
  journal={Physics Reports},
  volume={895},
  pages={1--28},
  year={2021},
  publisher={Elsevier}
}

@article{tretiakov2007vortices,
  title={Vortices in thin ferromagnetic films and the skyrmion number},
  author={Tretiakov, O. A and Tchernyshyov, O},
  journal={Physical Review B},
  volume={75},
  number={1},
  pages={012408},
  year={2007},
  publisher={APS}
}

@article{gobel2019magnetic,
  title={Magnetic bimerons as skyrmion analogues in in-plane magnets},
  author={G{\"o}bel, B{\"o}rge and Mook, Alexander and Henk, J{\"u}rgen and Mertig, Ingrid and Tretiakov, Oleg A},
  journal={Physical Review B},
  volume={99},
  number={6},
  pages={060407},
  year={2019},
  publisher={APS}
}

\end{document}

%\preprint{APS/123-QED}

\begin{center}
    {\fontsize{14}{17}\selectfont\bfseries Supplementary Material}

\vspace{0.4cm}

{\fontsize{14}{17}\selectfont
\bfseries
Unveiling Three-Dimensional Skyrmion Transitions Through Vortices and Monopoles
}
%\title{Unveiling Three-Dimensional Skyrmion Transitions Through Vortices and Monopoles}

\author{M. E. Henderson}
\email{hendersonme@ornl.gov}
\affiliation{Oak Ridge National Laboratory, Oak Ridge, TN 37831, USA}
\affiliation{Institute for Quantum Computing, University of Waterloo, Waterloo, ON, Canada, N2L3G1}
\affiliation{Department of Physics \& Astronomy, University of Waterloo,
  Waterloo, ON, Canada, N2L3G1}

\author{D. Kurebayashi}
\affiliation{School of Physics, The University of New South Wales, Sydney 2052, Australia}

\author{B. Heacock}
\affiliation{National Institute of Standards and Technology, Gaithersburg, Maryland 20899, USA}

\author{W. Chen}
\affiliation{National Institute of Standards and Technology, Gaithersburg, Maryland 20899, USA}

\author{C. W.  Clark}
\affiliation{National Institute of Standards and Technology, Gaithersburg, Maryland 20899, USA}

\author{D. G. Cory}
\affiliation{Institute for Quantum Computing, University of Waterloo, Waterloo, ON, Canada, N2L3G1}
\affiliation{Department of Chemistry, University of Waterloo, Waterloo, ON, Canada, N2L3G1}

\author{D. Sarenac}
\affiliation{Department of Physics, University at Buffalo, State University of New York, Buffalo, New York 14260, USA}

\author{S. Watson}
\affiliation{National Institute of Standards and Technology, Gaithersburg, Maryland 20899, USA}

\author{J. S. White}
\affiliation{Laboratory for Neutron Scattering and Imaging, PSI Center for Neutron and Muon Sciences, Villigen, Switzerland}

\author{L. DeBeer-Schmitt}
\email{debeerschmlm@ornl.gov}
\affiliation{Oak Ridge National Laboratory, Oak Ridge, TN 37831, USA}

\author{O. A. Tretiakov}
\email{o.tretiakov@unsw.edu.au} % removed the stray space
\affiliation{School of Physics, The University of New South Wales, Sydney 2052, Australia}

\author{D. A. Pushin}
\email{dmitry.pushin@uwaterloo.ca}
\affiliation{Institute for Quantum Computing, University of Waterloo, Waterloo, ON, Canada, N2L3G1}
\affiliation{Department of Physics \& Astronomy, University of Waterloo,
 Waterloo, ON, Canada, N2L3G1}

\date{\today}

\maketitle
\end{center}
% Optional (recommended) even for supplements:
% \begin{abstract}
% \end{abstract}

\suppsection{Neutron Scattering Tomography and Reconstruction}
Experimental reconstructions were generated via a small angle neutron scattering (SANS) tomography technique introduced in \cite{henderson2023three, heacock2020neutron}. Experimental inputs to the tomographic reconstruction were comprised of 3D multi-projection SANS measurements and magnetization parameters extracted from magnetic susceptibility measurements. In the former, two-dimensional SANS images were collected as a function of sample rotation---reminiscent of  traditional $I(q_x, q_y, \theta)$ rocking curve-type scans. Here, the entire experimental setup, consisting of both the sample and the external magnetic field, was rotated synchronously about two axes (rotation and tilt axes shown in Supplementary Figure \ref{fig:schematic}) . This procedure preserved the magnetic field orientation across tomographic rotations, thus ensuring a static internal magnetic configuration throughout the measurement sequence. In the latter, magnetic susceptibility measurements were performed using a Quantum Design MPMS 5 Superconducting Quantum Interference Device (SQUID) \footnote{Identification of a commercial product does not imply recommendation or endorsement by the National Institute of Standards and Technology, nor does it imply that the product is necessarily the best for the stated purpose} with an AC option installed to extract saturation magnetization and average magnetization values \cite{henderson2021characterization}. Preliminary SANS measurements were performed on the sample on the SANS-I instrument at the Paul Scherrer Institute (PSI). These measurements established the thermal equilibrium skyrmion phase diagram and measurement geometries used for the subsequent tomographic measurements taken on GP-SANS.

The sample was mounted on an aluminum plate in the Mag-G 11 T horizontal field magnet on the GP-SANS beamline at the High Flux Isotope Reactor at Oak Ridge National Laboratory \cite{wignall201240,heller2018suite}. The instrument was operated in its maximal low-q scattering configuration, with an upstream source-to-sample distance of $L_1 =  19 $ m and a downstream sample-to-detector distance of $L_2 = 19.56 $ m. Given the cube shape of the sample, a sample aperture of 4 mm diameter was chosen to maximize intensity and minimize sample flares. A triangular neutron wavelength distribution was used with $\Delta \lambda / \lambda \approx 0.13$, defined by a central wavelength of 10 \AA. These instrument parameters were consistent across all  measurements, selected to achieve the required q-space, given an average peak location of $q_{0}$ of 0.0052 \AA $^{-1}$. 

\begin{figure}[!t]
\centering
\includegraphics[width=\columnwidth]{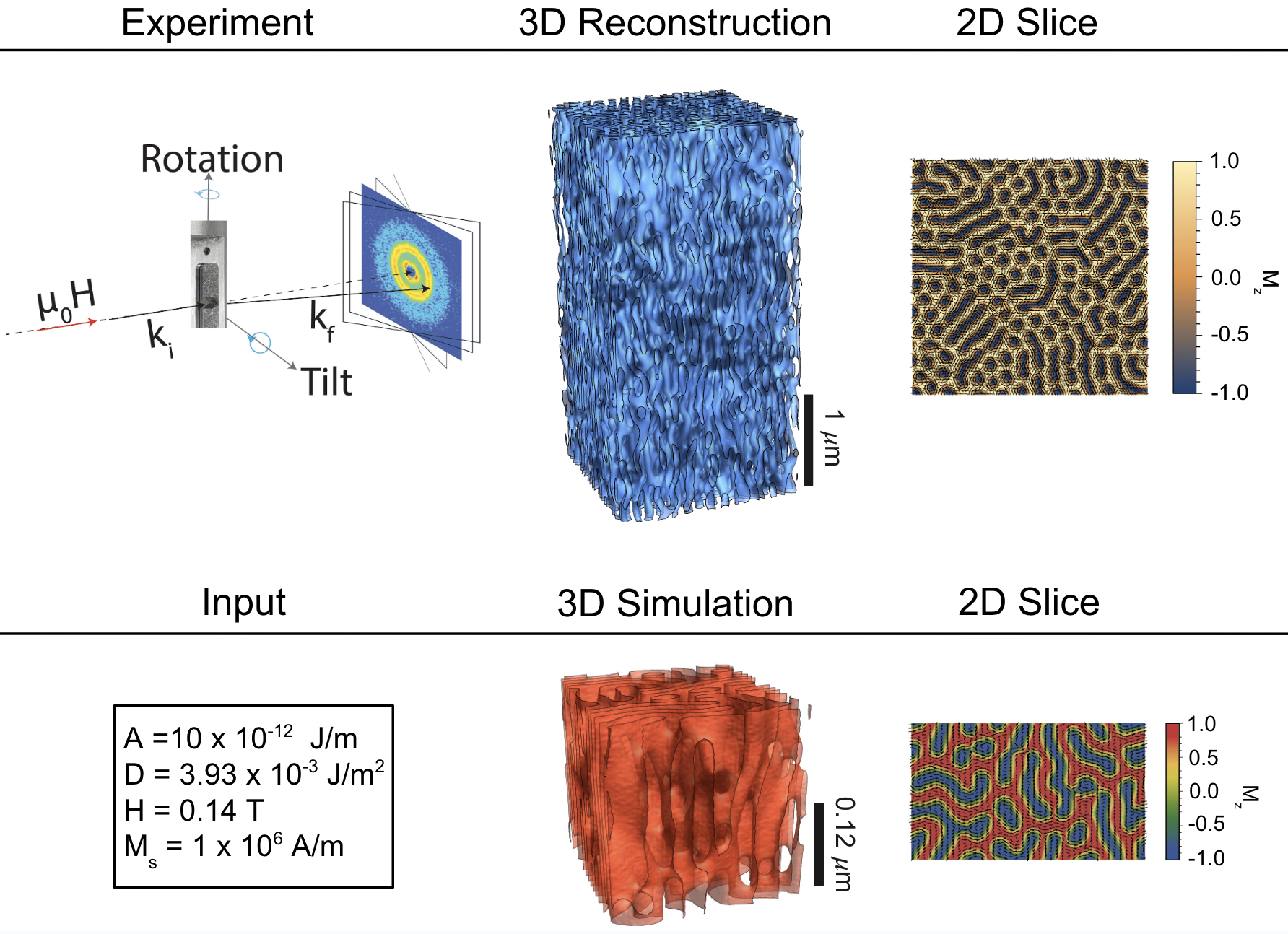}
\caption{Schematic of experimental reconstruction and simulation processes. Experimental multi-projection SANS datasets were collected via coincident rotation of the sample and magnetic field setup about vertical (rotation) and horizontal (tilt) axes. Monte Carlo simulations were performed with a randomly varying exchange field, bulk-type DMI, and Zeeman term, with constants A, D, and H outlined in the lower box, respectively. Magnetization volumes and in-plane magnetization slices are shown for the experimental tomographic reconstructions (upper plot) and supporting Monte Carlo simulations (lower plot). For both reconstruction and simulation magnetization volumes, the three-dimensional contours correspond to the z-component of the magnetization, $m_{z} = -0.5$.  }     

\label{fig:schematic}
\end{figure}

Empty measurements, with no sample mounted in the beam, were used to determine the beam center during data reduction processes. Transmission measurements were performed with the sample in either a paramagnetic state ($T > T_{\mathrm{C}}$, where $T_{\mathrm{C}}$ is the Curie temperature) or field-polarized state ($H > 0.6$ T), given an attenuation setting of x30. Rotation and tilt tomography angles spanned $\pm 10^\circ$ and $\pm 5^\circ$ in one degree increments, respectively. These ranges were chosen to capture the entire rocking curve peak and decay for each sample state. Angular step sizes were determined from rocking curve widths and structure, taking care to optimize resolution versus counting time. A constant temperature of 310 K was maintained throughout the entire experiment, corresponding to the maximal skyrmion formation temperature within the envelope of the thermal equilibrium phase. This temperature was previously identified through magnetic susceptibility measurements, and verified in-situ through measurements of neutron scattering intensity as a function of temperature. The magnetic field was swept from 0 T to 0.035 T, spanning the helical to ferromagnetic phase boundaries. Disordered states were achieved naturally upon field nucleation from the helical state, whereas an additional symmetry-breaking field rotation sequence was required to generate ordered skyrmion states. For these states, the sample was rotated sequentially in the magnetic field $\pm 90^\circ$ via an independent motor for sample stick rotation. This skyrmion ordering method was first developed in \cite{gilbert2019precipitating} and has been subsequently applied to this compositional series in \cite{henderson2021characterization,henderson2022skyrmion, henderson2023three}.

Additional experimental reconstruction inputs required for the reconstruction process were extracted from magnetic susceptibility measurements performed in \cite{henderson2021characterization, henderson2023three}. A saturation magnetization of $4\pi M_{s} = $ 1900 G was used for all reconstructions, while the Zeeman term, $h$, was tuned for each field-dependent reconstruction. The selected values of $h$ were chosen such that the average magnetization, $\langle m_z \rangle$, computed from the reconstruction matched those estimated from DC susceptibility measurements in \cite{henderson2021characterization}. Average magnetization values spanning the helical (0 T) to high-field skyrmion boundary (0.035 T) ranged from $\langle m_z \rangle =$ 0 to $\langle m_z \rangle = - 0.66 $. The scattering dimensions were set by the pixel dimensions of the 2D SANS detector,  $N$ x  $N = 128$ x $128$, while the calculated q-space resolution was $dq = 3.3$ $\mu m^{-1}$. From this, $dx$ was calculated according to $dx = 2\pi / (dqN)$, yielding 14.8 nm. In an effort to optimize the computation times of the reconstructions, the reconstruction height was set to $N'$ = 512 pixels. %The chosen height is reasonable given the correlation lengths of the measured skyrmion states, in combination with the angular ranges and resolutions of the tomographic datasets. 
Together, this yielded a reconstruction volume of $1.9$ $\mu m$ x $1.9$ $\mu m$ x $7.6$ $\mu m$.

\suppsection{Additional Topological Structures and Transition Pathways}

\begin{figure}[H]

\includegraphics[width =\textwidth]{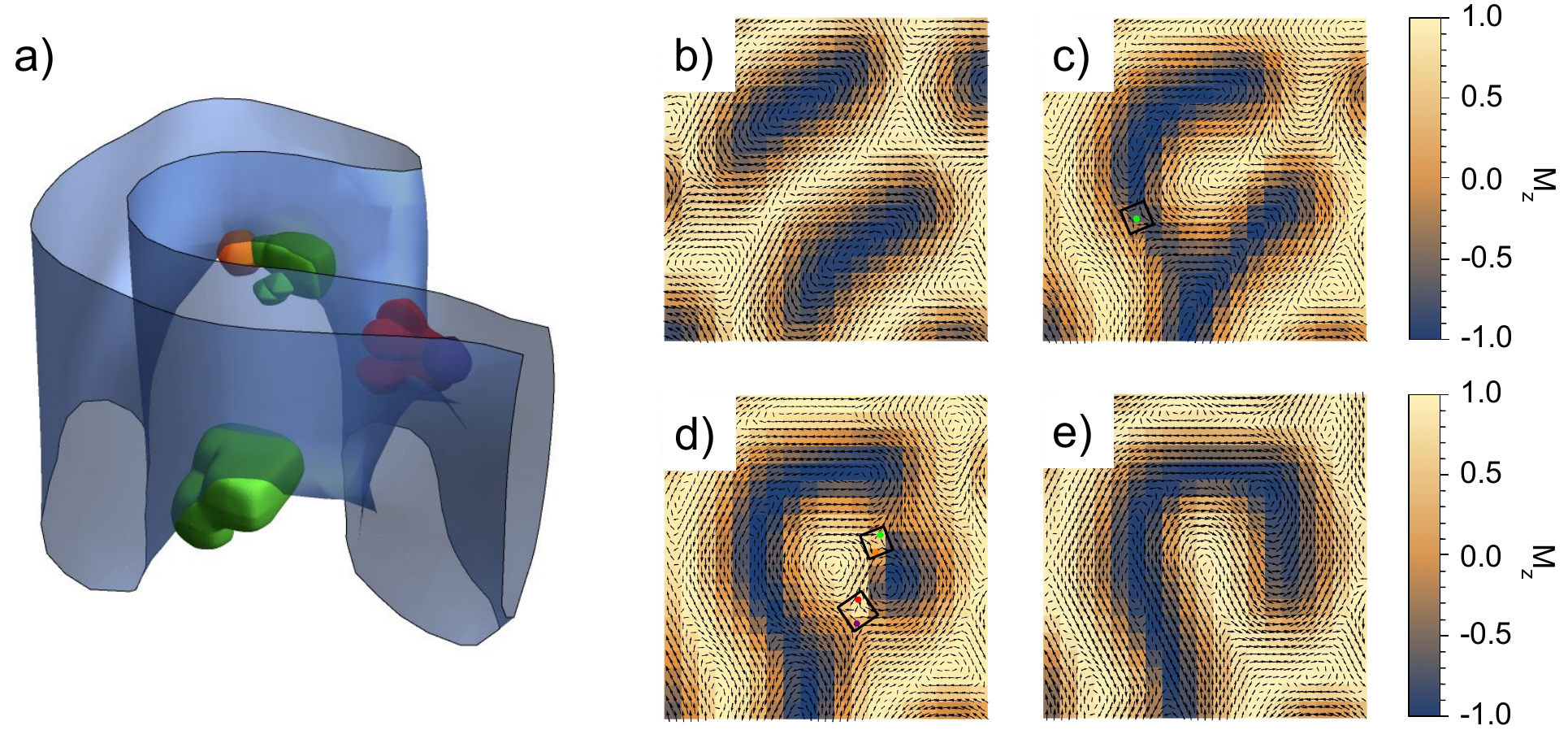}
\caption{Experimentally observed bound meron-antimeron topological transition in three dimensions. Three-dimensional contours (a) of the z-component of the magnetization, $m_{z} = -0.5$, and the emergent magnetic charge density, $\rho_{em}$, show a coupling of merons and S$^{-}$ (orange)/B$^{+}$ (green) and S$^{+}$ (purple)/B$^{-}$ (red) (anti)monopoles. In-plane magnetization slices (b-e) show the merging and dissociation of two skyrmions through merons. Black boxes highlight antivortex and meron topological transition points  with circular dots indicating monopole/antimonopole locations, colored according to their branching/segmenting genre.}        

\label{fig:meron_antimeron}
\end{figure}

Composite topological states comprised of bound merons with opposite $Q$ are observed in the disordered 0.025 T reconstructions (Supplementary Fig.~\ref{fig:meron_antimeron}). These states serve as an additional genre of hybrid $Q = 0$ skyrmion-based structure. Forming in a similar manner to skyrmionium states, the $Q = 0$ hybrid meron bundles exhibit vortex-antivortex transition pathways through MP/AMP pairs. In particular, two elongated skyrmions merge, trapping a vortex at their centre, with antivortices pinned to the merging locations. From this, the criticality of vortices and antivortices in stabilizing various exotic skyrmion-based $Q = 0$ structures becomes apparent.

\begin{figure}[!t]
\centering
\includegraphics[width=\textwidth]{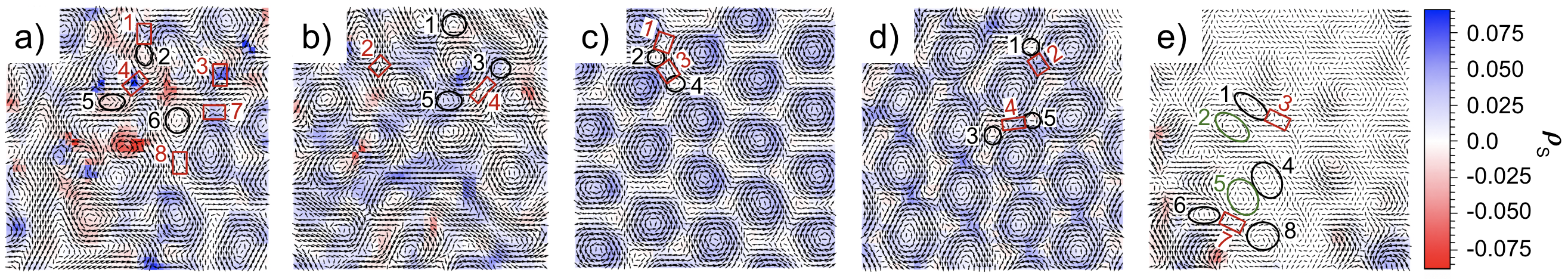}
\caption{ Coexisting skyrmion and vortex-antivortex lattice structures for field increasing experimental reconstructions. In-plane slices show the topological charge density for various skyrmion hosting fields given disordered states at 0.015 T (a) and 0.025 T (b), in addition to ordered states at 0.025 T (c) and 0.035 T (d-e). Black arrows represent the xy component of spin and $\rho_{s}$ denotes the calculated topological charge density. Selected regions containing vortices (black, green) and antivortices (red) are highlighted, with corresponding topological charges calculated in Supplementary Table \ref{tab:table}. Slice e) shows void regions where the skyrmions have completely annihilated, highlighting vortex structures in the background magnetization which may serve as remnants of vortex-mediated topological transitions.          
}
\label{fig:vortex}
\end{figure}

Vortex-antivortex lattices are observed throughout the entire skyrmion phase envelope, irrespective of disorder and field. Supplementary Fig.~\ref{fig:vortex} outlines in-plane magnetization slices and their topological charge densities for various disordered (D) and ordered (O) field-increasing reconstructions spanning 0.015 T - 0.035 T. Representative vortex and antivortex regions are highlighted, with their corresponding topological charges ($Q$) calculated in Supplementary Table \ref{tab:table}. While vortex structures are observed throughout all reconstructions within the skyrmion phase, it is immediately apparent that (1) the more ordered the state and (2) the closer the state is in field to its maximal skyrmion formation value, the lower the associated topological charge of the vortices. That is to say, reconstructions which are either disordered (Supplementary Fig.~\ref{fig:vortex}b) or lay close to the boundaries of the skyrmion phase window (Supplementary Fig.~\ref{fig:vortex}a, d) tend to exhibit vortices with significant non-trivial topological charge. These regions in phase space tend to be the most topologically active, undergoing nucleation events (in low-field conditions), annihilation events (in high-field conditions), and labyrinth/domain dissociation events (in disordered conditions). Non-trivial vortices may therefore be said to cluster in regions of phase-space where topology is changing.

The action of these topological transition dynamics through branching and segmenting events manifests distorted and elongated structures which effectively reduce the overall topological charge of the collective state. As a result, we note an inverse relationship between the total topological charge of the skyrmion state and the topological charge of the vortex structures within the state. This suggests that vortices play a critical role in mediating topological transformations within the skyrmion state, imparting and exchanging their topologies with skyrmions to facilitate field-induced transitions, after which they relax to trivial distortions of the background magnetization. The presence of vortices with vanishing topological charge in skyrmion-depleted void-like regions (as observed in Supplementary Fig.~\ref{fig:vortex}e) is further consistent with vortices acting as transient conduits for such topological exchanges. In particular, the residual vortices present in high-field reconstructions undergoing skyrmion annihilation events may serve as remnants/footprints for previous topologically active transition regions. Here, the trivial magnetization curls may represent the relaxed low-Q distortions of former pinning sites for 3D defects or frozen distortions from domain-wall fragments. While the magnetization tends towards a uniform field-polarized alignment with increasing field towards the ferromagnetic boundary, various factors such as pinning, disorder, and finite thickness may result in residual vortices with vanishing $Q$. The persistence of these trivial vortices, even in the absence of skyrmion topology, demonstrates that vortices may act as transient mediators of topology, reducing the nucleation energy barrier for 3D singular defects and providing the geometry through which they may propagate.                 

\begin{table}[!t]
\centering
\begin{tabular}{|c|c|c|c|c|}
\hline
\multicolumn{1}{|c|}{} 
& \multicolumn{1}{c|}{\makecell{\textbf{a} \\ (0.015 T, D)}} 
& \multicolumn{1}{c|}{\makecell{\textbf{b} \\ (0.025 T, D)}} 
& \multicolumn{1}{c|}{\makecell{\textbf{c} \\ (0.025 T, O)}} 
& \multicolumn{1}{c|}{\makecell{\textbf{d} \\ (0.035 T, O)}} \\
\hline
1 & 0.26 & 0.04 & 0.04 & 0 \\
2 & -0.15 & 0.08 & 0 & 0.32 \\
3 & 0.31 & 0.2 & 0.07 & 0.0 \\
4 & 0.3 & 0.03 & 0.02 & 0.38 \\
5 & -0.22 & 0.33 &  & 0 \\
6 & -0.13 &  &  &  \\
7 & 0.14 &  &  &  \\
8 & 0.05 &  &  &  \\
\hline
\end{tabular}
\caption{Topological charge for highlighted regions in plots a-d of Supplementary Fig.~\ref{fig:vortex}. Experimental magnetic field settings are listed for each selected 2D slice taken from the reconstructions, with ordered and disordered states denoted by (O) and (D). Plot e) is omitted as all calculated topological charge values are zero.}
\label{tab:table}
\end{table}

\bibliographystyle{apsrev4-2}
\bibliography{references}